\documentclass[conference]{IEEEtran}
\usepackage[latin9]{inputenc}
\usepackage{amsmath}
\usepackage{graphicx}
\usepackage{multirow}
\usepackage{multicol}
\usepackage{booktabs}
\usepackage{adjustbox}
\usepackage{xcolor}
\ifCLASSINFOpdf
\else
\fi

\usepackage{stfloats}
\usepackage{url}

\begin{document}
%
\title{Texture-based Presentation Attack Detection for Automatic Speaker Verification}

\author{\IEEEauthorblockN{Lazaro J. Gonzalez-Soler}
\IEEEauthorblockA{Biometrics and Internet Security Research Group \\ Hochschule Darmstadt, Germany \\
lazaro-janier.gonzalez-soler@h-da.de}
\and
\IEEEauthorblockN{Jose Patino}
\IEEEauthorblockA{EURECOM, France\\
jose.patino@eurecom.fr}
\and
\IEEEauthorblockN{Marta Gomez-Barrero}
\IEEEauthorblockA{Hochschule Ansbach, Germany \\
marta.gomez-barrero@hs-ansbach.de}
\and
\IEEEauthorblockN{Massimiliano Todisco}
\IEEEauthorblockA{EURECOM, France\\
massimiliano.todisco@eurecom.fr}
\and
\IEEEauthorblockN{Christoph Busch}
\IEEEauthorblockA{Biometrics and Internet Security Research Group \\ Hochschule Darmstadt, Germany \\
christoph.busch@h-da.de}
\and
\IEEEauthorblockN{Nicholas Evans}
\IEEEauthorblockA{EURECOM, France\\
nick.evans@eurecom.fr}}


%


\maketitle

\begin{abstract}
Biometric systems are nowadays employed across a broad range of applications.  They provide high security and efficiency and, in many cases, are user friendly. Despite these and other advantages, biometric systems in general and Automatic speaker verification (ASV) systems in particular can be vulnerable to attack presentations. The most recent ASVSpoof 2019 competition showed that most forms of attacks can be detected reliably with ensemble classifier-based presentation attack detection (PAD) approaches. These, though, depend fundamentally upon the complementarity of systems in the ensemble.
With the motivation to increase the generalisability of PAD solutions, this paper reports our exploration of texture descriptors applied to the analysis of speech spectrogram images. In particular, we propose a common fisher vector feature space based on a generative model. Experimental results show the soundness of our approach: at most, 16 in 100 bona fide presentations are rejected whereas only one in 100 attack presentations are accepted.            
\end{abstract}


%
\IEEEpeerreviewmaketitle

\section{Introduction}
\label{sec:intro}

In the last decades, biometric systems have experienced a broad development since they provide high security, efficiency, and in many cases they are more user friendly than the traditional credential-based access control systems. In spite of those advantages, they are vulnerable to attack presentations, which can be easily carried out by a non-authorised individual without having enough computational knowledge~\cite{ISO-IEC-30107-3-PAD-metrics-170227}. Those, in which biometric systems, are commonly deployed are in turn employed to gain access to several applications such as bank accounts, to unlock smartphones, or to circumvent a border control. Specifically, Automatic Speaker Verification (ASV) systems have shown a verification performance deterioration when different Presentation Attack Instruments (PAIs) such as replay, voice conversion, speech synthesis and impersonation are launched~\cite{Alegre-PAD-LBPOnC-2013}.

In order to deal with those security threats, several Presentation Attack Detection (PAD) approaches have been proposed. These methods try to determine whether a given sample stems from a live subject (i.e., it is a bona fide presentation, BP) or from a synthetic replica (i.e., it is an attack presentation, AP). According to the last ASVSpoof 2019 competition~\cite{Todisco-ASVSpoof-arXiv-2019}, the top Deep Neural Networks (DNN)-based PAD techniques have achieved a remarkable detection performance to spot PAI species which can be either known or unknown a priori in training time. In spite of the promising results reported, those methods are time-consuming, hence being unsuitable for lightweight security applications, e.g. for smartphones.         

To overcome the aforementioned drawbacks, very few works have explored texture-based representation for audio PAD. Alegre \textit{et al.}~\cite{Alegre-PAD-LBPOnC-2013} proposed a PAD method based on the combination of Local Binary Patterns (LBP) and one-class classifiers. Even if the proposed algorithm reported a poor detection performance for some unknown attacks such as voice conversion, this work showed the generalisation capability of the proposed texture-based representation for audio PAD. Motivated by that fact, we explore in this work several image processing texture descriptors in combination with a Support Vector Machine (SVM), which have been successfully employed for fingerprint~\cite{Marasco-fPAD-Survey-2014} and face~\cite{Galbally-facePAD-Survey-2014} PAD. SVMs are popular as classifiers since they perform well in high-dimensional spaces and avoid over-fitting. In order to improve the generalisation capability of the analysed texture descriptors, we also utilise the Fisher Vector (FV) representation, which has shown remarkable results for unknown fingerprint~\cite{Gonzalez-PAD-MaterialImpact-ICB-2019} and face~\cite{Gonzalez-Soler-PAD-FvEnc-BIOSIG-2020} PAD. By assuming that the unknown attacks share more texture, shape, and appearance features with known PAI species than with BP samples, the FV representation defines a common feature space from the parameters learned by an unsupervised Gaussian Mixture Model (GMM) in order to deal with the generalisation to unknown attacks.  

The remainder of this paper is organised as follows. The speech-to-image domain transformation techniques are presented in Sect.~\ref{sec:transformation}. Sect.~\ref{sec:proposal} briefly describes the texture descriptors analysed in this work. Sect.~\ref{sec:experiments} describes the experimental protocol and Sect.~\ref{sec:results} presents the results. Finally, conclusions and future work directions are presented in Sect.~\ref{sec:conclusion}.  

\section{From speech to images domain}
\label{sec:transformation}  

\begin{figure}[!t]
	\centering
	\includegraphics[width=0.9\linewidth]{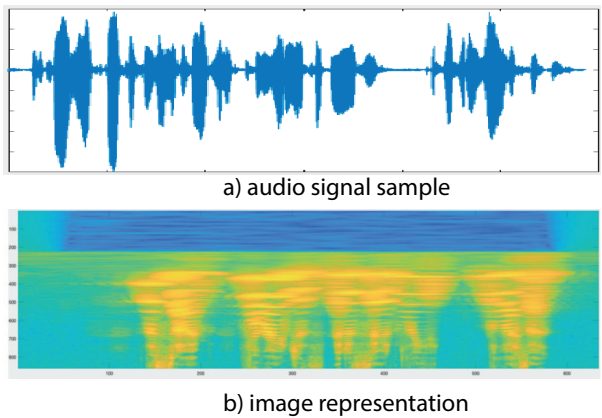}
	\caption{A speech sample together with its texture image representation.}
	\label{fig:signal_image}
\end{figure} 

The visualisation of audio/speech signals is key to many audio analysis tasks, usually involving: (i)~time-domain, (ii)~frequency-domain, or (iii)~time-frequency-domain representations known as spectrograms, which show the signal amplitude over time at a set of discrete frequencies.
Many time-frequency representations have been proposed, each with different characteristics. Keeping in mind that an audio signal can be represented as an image, as shown in Fig.~\ref{fig:signal_image}, in this work we focus on the following four time-frequency representations: (i)~short-time Fourier transform (STFT)~\cite{Oppenheim-SignalProc-99}, (ii)~linear frequency cepstral coefficients (LFCC)~\cite{Tak-IS-2020}, (iii)~constant Q transform (CQT)~\cite{Brown-CQCC-91}, and (iv)~constant Q cepstral coefficients (CQCC)~\cite{Todisco-PAD-CQCC-2017,Tak-OD-2020}.

The STFT is a time-frequency decomposition based upon the application of Fourier analysis to short segments or windows of the audio signal. As such, it is effectively a filter bank where the bandwidth of each filter is constant and is related to the window function. 

LFCC coefficients are computed from the STFT by applying the discrete cosine transform (DCT)~\cite{Oppenheim-SignalProc-99}. Generally, only lower-order coefficients are retained since they represent the vocal tract configuration.

The CQT is a perceptually motivated approach to time-frequency analysis. In contrast to Fourier-based approaches, the bin frequencies of the filterbank are geometrically distributed. Compared to the STFT, the CQT has a greater frequency resolution for lower frequencies and a greater temporal resolution for higher frequencies.

CQCCs stem from the application of cepstral processing to CQT representations. CQCCs offer a time-frequency resolution more closely related to that of human perception. These features were designed specifically for PAD but have also shown to be beneficial for ASV and utterance verification~\cite{Todisco-ASVIS-2016}.

\section{Texture analysis based approaches}
\label{sec:proposal}

In order to analyse the texture features extracted from the time-frequency image representations described in Sect.~\ref{sec:transformation}, we selected several well-known texture descriptors, which have been successfully employed for the PAD task across different modalities such as fingerprint~\cite{Marasco-fPAD-Survey-2014} and face~\cite{Galbally-facePAD-Survey-2014}: Local Binary Patterns (LBP)~\cite{Ojala-LBP-PAMI-2002}, Multi-block LBP (MB-LBP)~\cite{Zhang-MB-LBP-ICB-2007}, Local Phase Quantisation (LPQ)~\cite{Ojansivu-LPQ-2008}, and Binarized Statistical Image Features (BSIF)~\cite{Kannala-BISF-ICPR-2012}.

\subsection{Local Binary Patterns}
\label{sec:LBP}

LBP~\cite{Ojala-LBP-PAMI-2002} represents an image as a histogram of uniform patterns which allows capturing shape and texture features from the image. Specifically, given a circular image patch $X$ with radii $\sigma$ and $N$ pixels around the center, the LBP descriptor is defined over $X$ as:
\begin{equation}
\mathrm{LBP}_{N, \sigma} = \sum_{i = 0}^{N - 1} f(g_i - g_c)2^i,
\end{equation}  

\noindent where $g_i$ with $i = 0 \ldots N - 1$ are gray intensity values around the center $g_c$. In addition, $f(g_i - g_c)$ is computed as:
\begin{align}
f(g_i - g_c) = \left\{ \begin{array}{cc} 
1, & \hspace{2mm} g_i - g_c \geq 0 \\
0, & \hspace{2mm} g_i - g_c  <  0 \\
\end{array} \right.
\end{align}

In order to capture more information and thereby increase the descriptor distinctiveness, we compute the LBP patterns for various radii $\sigma$ and number of neighbors $N$ (i.e., $\sigma$ = \{1, 2, 3\} and $N$ = \{8, 16, 24\}), as in~\cite{Gonzalez-PAD-MaterialImpact-ICB-2019}.

\subsection{Multi-block Local Binary Patterns}
\label{sec:MB-LBP}

The MB-LBP~\cite{Zhang-MB-LBP-ICB-2007} encodes the intensities of rectangular regions with the LBP operator, which allows describing several local structures of an image. Whereas the LBP descriptor in \cite{Ojala-LBP-PAMI-2002} is defined for each pixel by thresholding the 3 $\times$ 3 neighbourhood pixel values with the center pixel value, the MB-LBP operator represents each pixel $x$ by comparing the central rectangle average intensity $g_x$ with those of its neighbourhood rectangles. Therefore, it can detect numerous image structures such as lines, edges, spots, flat areas, and corners \cite{Zhang-MB-LBP-ICB-2007}, at different scales and locations. Unlike LBP, the MB-LBP descriptor can thus capture large scale structures that may be the dominant features of images, with 256 binary patterns. In our work, we compute the MB-LBP descriptor for several rectangle sizes $R_x$ = \{3, 5, 7, 9\}.     

\subsection{Local Phase Quantization}
\label{sec:LPQ}

LPQ~\cite{Ojansivu-LPQ-2008} is a texture descriptor designed to deal with blurred images. It represents an image patch of size $l \times l$ centred on a pixel $x$ as a 256-histogram by using the local phase information, extracted by a STFT. Let $F_{u_{i = 1\ldots 4}}$ be the outputs of the STFT for the pixel $x$ using four bi-dimensional spatial frequency $u_0$, $u_1$, $u_2$ and $u_3$, the LPQ features for $x$ are defined as a vector whose components are formed by stacking the real and imaginary part of $F_{u_{i = 1\ldots 4}}$. Subsequently, the vector elements are quantized using a previously defined function and then represented as a integer value in the range $[0\ldots 255]$. In order to make the LPQ coefficients statistically independent, a decorrelation step based on whitening transform was performed.        

\subsection{Binarized Statistical Image Features}
\label{sec:BSIF}

BSIF~\cite{Kannala-BISF-ICPR-2012} is a local descriptor based on Independent Component Analysis (ICA), which uses pre-learned filters to obtain statistically meaningful representation of the data. Let $X$ be an image patch of size $l \times l$ and $W = \lbrace W_1, \dots, W_N \rbrace$ a set of linear filters of the same size as $X$. Then, we can compute the binarised responses $b_i$, with $i = 1, \dots, N$, as:
\begin{align}
b_i = \left\{ \begin{array}{cc} 
1 & \hspace{5mm} \sum_{u,v}W_i(u,v)X(u,v) > 0 \\
0 & \hspace{5mm} \text{otherwise} \\
\end{array} \right.
\end{align}

All the filter responses $b_i$ are subsequently stacked to form a single bit string $b$ of size $N$ per image pixel. Consequently, that bit string is transformed in a decimal value, and then a $2^N$ histogram for $X$ is computed. This histogram is finally represented as a 128-component vector~\cite{Gonzalez-Soler-PAD-FvEnc-BIOSIG-2020}. In our work, we adopt 60 filter sets~\cite{Kannala-BISF-ICPR-2012} with different sizes $l$ = \{3, 5, 7, 9, 11, 13, 15, 17\} and number of filters $N$ = \{5, 6, 7, 8, 9, 10, 11, 12\} to compute the BSIF descriptors.  

\subsection{Fisher Vector encoding}
\label{sec:FV}

For FV encoding, we build upon the idea proposed by~\cite{Gonzalez-PAD-FVLocalSIFTEnc-ArXiv-2019,Perronnin-FVimproving-ECCV-2010}: a Gaussian Mixture Model (GMM) is used to unsupervisely learn embedding features from a texture descriptor. Generally, a GMM model based on $K$-components, which are represented by their mixture weights~($\pi_k$), Gaussian means~($\mu_k$), and covariance matrix~($\sigma_k$), with $k = 1, \dots, K$, allows discovering semantic sub-groups from known PAI and BP samples, which could successfully generalise to and hence enhance the detection of unknown PAI species. Given the parameters learned by a GMM model, the FV computes the average first-order and second-order statistics differences between the texture features and each GMM Gaussian, thereby yielding a $2Kd$ dimensional vector ($d$ is the texture feature dimension). The FV gradient computation between texture features extracted from an given unknown sample and the GMM components represents how its distribution differs from the distribution of the model. Therefore, a better data distribution will reveal more discriminative FV common feature space to detect unknown attacks.

\section{Experimental evaluation}
\label{sec:experiments}

The experimental evaluation follows a threefold objective: $i)$ analyse the detection performance of the texture descriptors for known and unknown PAD; $ii)$ establish a benchmark against the speech-based state-of-the-art PAD techniques; and $iii)$ evaluate the computational cost of our proposal.     

\subsection{Databases}
\label{sec:databases}

      


The experimental evaluation was conducted over the freely available ASVSpoof 2019 database~\cite{Todisco-ASVSpoof-arXiv-2019,WANG-ASVSpoof2019-2020}, which comprises two assessment scenarios: Logical Access (LAc) and Physical Access (PAc). Both LAc and PAc databases are partitioned in three disjoint datasets: training, development, and evaluation. Whereas the PAIs in the training and development datasets were built with the same algorithms and capture conditions (i.e., it is the known attack scenario), PAIs for the evaluation dataset were generated with different techniques and capture conditions (i.e., it is the unknown attack scenario).   

The LAc partition contains PAI samples which were generated using 17 different text-to-speech synthesis (TTS) and voice conversion (VC) technologies: six were designated for known attack assessment (i.e., A01-A06) and 11 for unknown attack (i.e., A07-019 with exception of A16 = A04 and A19 = A06 and hence both two attacks are in the training set). 
In order to analyse and improve the ASV reliability in different acoustic environments and replay setups, the training and development data for the PAc scenario is created under 27 different acoustic and 9 replay configurations. The replay settings comprise 3 attacker-to-talker (i.e., A, B, C) recording distances and 3 loudspeaker quality (i.e., A, B, C). The evaluation dataset is generated in the same manner as training and development data but with different random acoustic and replay configurations.

\subsection{Baselines and metrics}
\label{sec:metrics}

In order to establish a fair benchmark, we adopt two PAD baseline approaches, which use GMM back-end classifier with either CQCC ($B01$) or LFCC ($B02$) features. It should be noted that whereas the baseline approaches employed a bi-cluster GMM model for a BP or AP classification, our analysed FV approach uses it as generative model to fit the BP and AP data distribution. Based on that fact, we evaluated several numbers of Gaussian clusters (i.e., $K$ = \{64, 128, 256, 512, 1024\}) for the FV representation. 

The experimental evaluation is conducted in compliance with ISO/IEC 30107-Part 3~\cite{ISO-IEC-30107-3-PAD-metrics-170227}, analysing two metrics: $i)$ Attack Presentation Classification Error Rate (APCER) which is defined as the proportion of APs wrongly classified as BPs; and $ii)$ Bona Fide Presentation Classification Error Rate (BPCER): which is defined as the proportion of BPs misclassified as APs. In addition, we report the Detection Error Trade-off (DET) curves between both errors, the BPCER values for several security thresholds (i.e., BPCER10, BPCER20, BPCER100), and the D-EER. Finally, to evaluate the performance between the proposed PAD approaches and a ASV system, we report the minimum normalised tandem detection cost function (t-DCF)~\cite{Kinnunen-tDCF_2020}, which is the primary metric used for the ASVspoof 2019 challenge\footnote{\url{https://www.asvspoof.org/}}.



\section{Results and discussion}
\label{sec:results}

\subsection{Known attacks}

In order to analyse the detection performance, two sets of experiments were carried out. In the first experiment set, we optimise the detection performance of our texture descriptors in terms of the D-EER for different parameter configurations. Table~\ref{tab:known-attacks} shows the D-EER for the best parameter setting over the development set in the LAc and PAc scenarios. As it may be observed, among all speech-to-image domain transformations, the CQT reports the best detection performance, thereby resulting a mean D-EER of 6.14\% and 5.80\% for the LAc and PAc scenarios, respectively. In addition, among the texture descriptors, the BSIF unveils the best texture features for the audio PAD task: D-EERs of 0.86\% and 4.54\% are attained for the LAc with STFT and PAc with CQT, respectively, thereby showing its suitability for the audio PAD task.

\begin{table}[!t]
	\begin{center}
    \caption{Benchmark in terms of D-EER(\%) of the texture descriptors for the best parameter configuration per speech-to-image domain transformation for known attacks.}\label{tab:known-attacks} 
    \vspace{-.5cm}
    \setlength\tabcolsep{1.8pt}
    \begin{tabular}{lcccccccc} \toprule
                       & \multicolumn{2}{c}{\textbf{CQCC}}	& \multicolumn{2}{c}{\textbf{LFCC}} & \multicolumn{2}{c}{\textbf{STFT}} & \multicolumn{2}{c}{\textbf{CQT}}				\\ 
                       &      LAc          &      PAc         &     LAc          &     PAc            &     LAc           &      PAc           &    LAc        &  PAc	             \\   \midrule 
   Best BSIF           & \textbf{$N$ = 6} & \textbf{$N$ = 8}& \textbf{$N$ = 9}& \textbf{$N$ = 11} & \textbf{$N$ = 12} & \textbf{$N$ = 9}  & \textbf{$N$ = 9}   & \textbf{$N$ = 9}   	 \\
    Parameters         & \textbf{$l$ = 17}& \textbf{$l$ = 3}& \textbf{$l$ = 5}& \textbf{$l$ = 7}  & \textbf{$l$ = 15} & \textbf{$l$ = 5}  & \textbf{$l$ = 17}  & \textbf{$l$ = 13} 		\\  \midrule
      LBP              & 32.72            & 16.39           &    17.39        &    28.42          &   10.12           &   15.50           &  9.77        &   7.65          \\
      LPQ              & 20.72            & 13.80           &    19.20        &    43.59          &   12.66           &   15.04           &  9.54        &   6.05           \\
      BSIF             & \textbf{18.53}   & \textbf{13.11}  & \textbf{14.30}  & \textbf{18.08}    & \textbf{0.86}     &  \textbf{11.30}   &\textbf{2.11} & \textbf{4.54}   \\
      MB-LBP           & 19.68            & 22.30           &    16.52        &    24.28          &   1.10            &   12.01           &  3.12        &   4.98           \\          
      avg              & 22.91            & 16.40           &    16.85        &    28.59          &   6.19            &   13.46           &  6.14        &   5.80            \\ \bottomrule
			
		\end{tabular}
	\end{center}
	\vspace*{-0.5cm}
\end{table} 

\begin{table}[!t]
	\begin{center}
    \caption{Benchmark of our FV method and BSIF for known attacks.}\label{tab:fv-known-attacks-LA-PA}  
    \vspace{-.5cm} 
    \setlength\tabcolsep{2.1pt}
    \begin{tabular}{clcccccccc} \toprule
                     &  & \multicolumn{2}{c}{\textbf{CQCC}}	& \multicolumn{2}{c}{\textbf{LFCC}}  & \multicolumn{2}{c}{\textbf{STFT}}  & \multicolumn{2}{c}{\textbf{CQT}}        \\
                    &   & \multicolumn{2}{c}{$K$ = 512}     	& \multicolumn{2}{c}{$K$ = 512}      & \multicolumn{2}{c}{$K$ = 256}     & \multicolumn{2}{c}{$K$ = 512}  \\ \midrule  
      &\textbf{PAI}     &      D-EER     &  t-DCF          &     D-EER     &  t-DCF        &    D-EER       &  t-DCF        &  D-EER      &  t-DCF           			\\ \midrule 
      \multirow{7}{*}{\rotatebox{90}{Logical Access}}
      &A01              &       9.69     &   0.2973        &     5.70      &   0.1749      & \textbf{0.38}  &\textbf{0.0101}&   1.15      &  0.0454                 \\
      &A02              &       3.99     &   0.1202        &    11.22      &   0.3195      & \textbf{0.62}  &\textbf{0.0201}&   0.92      &  0.0294                 \\
      &A03              &      11.40     &   0.3503        &     6.59      &   0.2065      & \textbf{0.74}  &\textbf{0.0225}&   1.19      &  0.0388               \\
      &A04              &       8.76     &   0.2650        &    19.34      &   0.5258      & \textbf{0.84}  &\textbf{0.0282}&   2.68      &  0.0804            \\
      &A05              &       7.14     &   0.2260        &    11.40      &   0.3374      & \textbf{1.50}  &\textbf{0.0516}&   2.46      &  0.0715           \\
      &A06              &      17.77     &   0.5044        &    12.73      &   0.5850      & \textbf{0.82}  &\textbf{0.0278}&   4.09      &  0.1331           \\
      &pooled           &      10.28     &   0.3060        &    13.27      &   0.3660      & \textbf{0.86}  &\textbf{0.0294}&   2.55      &  0.0748             \\ \midrule
     & & \multicolumn{2}{c}{$K$ = 256}     	& \multicolumn{2}{c}{$K$ = 128}      & \multicolumn{2}{c}{$K$ = 512}     & \multicolumn{2}{c}{$K$ = 256}  \\ \midrule
      \multirow{10}{*}{\rotatebox{90}{Physical Access}}  
      &AA               &    16.26       &   0.4332        &     39.64     &   0.9045      &    21.32   &   0.5416      &\textbf{8.11}& \textbf{0.2038}          \\
      &AB               &     7.65       &   0.2184        &     27.26     &   0.6930      &    14.28   &   0.3763      &\textbf{2.43}& \textbf{0.0702}         \\
      &AC               &     6.23       &   0.1743        &     22.34     &   0.5767      &     9.75   &   0.2530      &\textbf{2.30}& \textbf{0.0671}        \\
      &BA               &    15.09       &   0.3859        &     33.26     &   0.8107      &     9.54   &   0.2334      &\textbf{5.19}& \textbf{0.1291}         \\
      &BB               &     6.53       &   0.1814        &     23.05     &   0.5933      &     6.19   &   0.1568      &\textbf{1.45}& \textbf{0.0373}        \\
      &BC               &     5.45       &   0.1502        &     19.03     &   0.5069      &     4.64   &   0.1141      &\textbf{1.16}& \textbf{0.0352}         \\
      &CA               &    15.56       &   0.4072        &     32.51     &   0.7987      &     9.19   &   0.2299      &\textbf{5.25}& \textbf{0.1365}        \\
      &CB               &     6.42       &   0.1719        &     22.73     &   0.6009      &     6.20   &   0.1566      &\textbf{1.21}& \textbf{0.0341}        \\
      &CC               &     5.11       &   0.1374        &     19.23     &   0.5136      &     4.19   &   0.1105      &\textbf{0.96}& \textbf{0.0275}        \\
      &pooled           &     9.94       &   0.2675        &     27.34     &   0.6784      &     11.05  &   0.2700      &\textbf{3.68}& \textbf{0.0976}         \\ \bottomrule
    \end{tabular}
	\end{center}
	\vspace*{-0.5cm}
\end{table} 

In a second set of experiments, we evaluate our FV approach for the best texture descriptor (i.e., BSIF) for known attack detection (dev set). Table~\ref{tab:fv-known-attacks-LA-PA} shows the FV detection performance for the best number of Gaussian clusters per speech-to-image domain transformation. We can first observe that STFT achieves the same detection performance for the LAc scenario for the pooled database (i.e., all PAI species) than the one reported by the single BSIF descriptor: a D-EER of 0.86\%, which is approximately three times lower than the one attained by the FV encoding with the images in the CQT domain. Alternatively, a different detection performance is reported for the PAc scenario where a D-EER of 3.68\% for the CQT outperforms the one attained by STFT (i.e., 11.05\%). Finally, it may be noted that the t-DCF values for both the STFT on LAc and CQT on PAc are respectively below 0.05 and 0.14, hence indicating that the FV provides a high security against PAIs to the ASV systems.

\subsection{Unknown attacks}
\label{sec:unknown-attacks}

\begin{table}[!t]
	\begin{center}
    \caption{Benchmark in terms of D-EER(\%) of the texture descriptors for unknown attack detection.}\label{tab:unknown-attacks} 
    \vspace{-.1cm}
    \setlength\tabcolsep{4pt}
    \begin{tabular}{lcccccccc} \toprule
                       & \multicolumn{2}{c}{\textbf{CQCC}}	& \multicolumn{2}{c}{\textbf{LFCC}} & \multicolumn{2}{c}{\textbf{STFT}} & \multicolumn{2}{c}{\textbf{CQT}}				\\ \midrule 
      \textbf{Method}  &  LAc            &  PAc             &     LAc        &     PAc        &     LAc         &      PAc         &      LAc       &  PAc	 			\\ \midrule 
      LBP              & 24.61          & 17.87           &    28.30      &    25.88      &   22.44        &    16.26        &     17.43     &  7.16          \\
      LPQ              & 25.10          & 18.83           &    24.10      &    29.77      &   20.06        &    16.94        &     16.89     &  5.65           \\
      BSIF             & \textbf{20.35} & \textbf{14.29}  & \textbf{18.36}& \textbf{19.74}&   15.33        & \textbf{11.29}  &\textbf{14.73} & \textbf{4.94}   \\
      MB-LBP           & 24.84          & 28.27           &    20.11      &    25.20      & \textbf{10.69} &    13.26        &     15.48     &  6.13          \\          
      avg              & 23.73          & 19.82           &    22.72      &    25.15      &   17.27        &    14.44        &     16.13     &  5.97          \\ \bottomrule
			
		\end{tabular}
	\end{center}
\end{table} 

\begin{table*}[!t]
	\begin{center}
    \caption{Benchmark with the state of the art ($B01$ and $B02$) of our FV method and BSIF for unknown attacks.}\label{tab:fv-unknown-attacks-LA-PA} 
    \vspace*{-.1cm}
    \begin{tabular}{clcccccccccccc} \toprule
                    &   & \multicolumn{2}{c}{\textbf{CQCC}}	& \multicolumn{2}{c}{\textbf{LFCC}}  & \multicolumn{2}{c}{\textbf{STFT}}  & \multicolumn{2}{c}{\textbf{CQT}} &  \multicolumn{2}{c}{\textbf{B01}}   & \multicolumn{2}{c}{\textbf{B02}}     \\
                    &   & \multicolumn{2}{c}{$K$ = 512}     	& \multicolumn{2}{c}{$K$ = 512}      & \multicolumn{2}{c}{$K$ = 256}     & \multicolumn{2}{c}{$K$ = 512}    &  \multicolumn{2}{c}{$K$ = 2}  &   \multicolumn{2}{c}{$K$ = 2}   \\ \midrule 
    &  \textbf{PAI}     &      D-EER     &     t-DCF       &     D-EER     &  t-DCF        &    D-EER         &  t-DCF        &  D-EER        &      t-DCF     &    D-EER         &  t-DCF        &  D-EER        &  t-DCF       			\\ \midrule 
    \multirow{12}{*}{\rotatebox{90}{Logical Access}} 
      &A07              &       7.80     &    0.2445       &    17.78      &   0.5147      &    0.31          &   0.0100      &  4.45         &    0.1376      & \textbf{0.00}    &\textbf{0.0000}&  12.86        &  0.3263             \\
      &A08              &       6.62     &    0.1958       &     1.75      &   0.0466      &    1.25          &   0.0356      &  4.60         &    0.1431      & \textbf{0.04}    &\textbf{0.0007}&  0.37         &  0.0086             \\
      &A09              &       3.23     &    0.0944       &     1.16      &   0.0351      &    0.30          &   0.0089      &  0.89         &    0.0270      &    0.14          &   0.0060      &\textbf{0.00}  &  \textbf{0.0000}      \\
      &A10              &       9.29     &    0.2784       &    17.35      &   0.5111      &   10.81          &   0.3140      & \textbf{7.91} &\textbf{0.2438} &   15.16          &   0.4149      & 18.97         &  0.5089             \\
      &A11              &       2.18     &    0.0685       &     9.32      &   0.2688      &    1.40          &   0.0430      &  3.41         &    0.1032      & \textbf{0.08}    &\textbf{0.0020}&  0.12         &  0.0027              \\
      &A12              &       7.08     &    0.2212       &    17.21      &   0.4900      &    5.84          &   0.1689      &  5.08         &    0.1560      & \textbf{4.74}    &\textbf{0.1160}&  4.92         &  0.1197              \\
      &A13              &       8.30     &    0.2648       &    23.91      &   0.6964      &    5.01          &   0.1415      & \textbf{1.94} & \textbf{0.0634}&   26.15          &   0.6729      &  9.57         &  0.2519              \\
      &A14              &       8.83     &    0.2686       &     8.74      &   0.2585      &    3.15          &   0.0971      &  2.05         &    0.0638      &   10.85          &   0.2629      & \textbf{1.22} &  \textbf{0.0314}    \\
      &A15              &       4.56     &    0.1415       &     5.13      &   0.1517      &    4.32          &   0.1345      &  4.48         &    0.1377      & \textbf{1.26}    &\textbf{0.0344}&  2.22         &  0.0607             \\
      &A16              &       7.54     &    0.2363       &    15.22      &   0.4259      &    0.74          &   0.0234      &  2.04         &    0.0669      & \textbf{0.00}    &\textbf{0.0000}&  6.31         &  0.1419              \\
      &A17              &      34.39     &    0.9115       &    21.22      &   0.5745      &   31.20          &   0.8643      & 16.62         &    0.4766      &   19.62          &   0.9820      & \textbf{7.71} & \textbf{0.4050}       \\
      &A18              &      36.08     &    0.9536       &    32.19      &   0.8853      &    5.94          &\textbf{0.1793}& 10.28         &    0.3080      &    3.81          &   0.2818      & \textbf{3.58} &  0.2387               \\
      &A19              &      26.94     &    0.7321       &    23.08      &   0.6628      &    4.62          &   0.1388      & 11.49         &    0.3454      & \textbf{0.04}    &\textbf{0.0014}& 13.94         &  0.4635               \\
      &pooled           &      14.62     &    0.3691       &    16.79      &   0.3837      &    7.76          &\textbf{0.1881}& \textbf{6.83} &    0.1926      &    9.57          &   0.2366      &  8.09         &  0.2116               \\ \midrule
      && \multicolumn{2}{c}{$K$ = 256}     	& \multicolumn{2}{c}{$K$ = 128}      & \multicolumn{2}{c}{$K$ = 512}      & \multicolumn{2}{c}{$K$ = 256}    &     \multicolumn{2}{c}{$K$ = 2}  &   \multicolumn{2}{c}{$K$ = 2}         \\ \midrule 
      \multirow{10}{*}{\rotatebox{90}{Physical Access}}
      &AA               &    23.14       &  0.5396         &     37.94     &  0.8827       &    19.45   &   0.4954      & \textbf{7.64}& \textbf{0.1985} &    25.28       &      0.4975      &      32.48       &   0.7359    \\
      &AB               &    17.27       &  0.4162         &     26.89     &  0.7049       &    14.44   &   0.3714      & \textbf{2.05}& \textbf{0.0548} &    6.16        &      0.1751      &       4.40       &   0.1295    \\
      &AC               &    11.28       &  0.2859         &     22.55     &  0.5913       &    10.70   &   0.2772      &    2.17      &   0.0594        & \textbf{2.13}  & \textbf{0.0529}  &       3.95       &   0.1121    \\
      &BA               &    19.74       &  0.4967         &     31.24     &  0.7644       &    10.78   &   0.2785      & \textbf{5.17}& \textbf{0.1360} &   21.87        &      0.4658      &      24.59       &   0.6011     \\
      &BB               &    14.40       &  0.3679         &     23.11     &  0.6132       &     7.06   &   0.1877      & \textbf{1.22}& \textbf{0.0341} &   5.26         &      0.1483      &       4.29       &   0.1252    \\
      &BC               &     9.66       &  0.2517         &     19.32     &  0.5157       &     5.42   &   0.1458      & \textbf{1.23}& \textbf{0.0336} &   1.61         &      0.0433      &       3.20       &   0.0888    \\
      &CA               &    18.64       &  0.4709         &     28.35     &  0.7084       &     9.88   &   0.2568      & \textbf{5.46}& \textbf{0.1408} &  21.10         &      0.5025      &      21.63       &   0.5524    \\
      &CB               &    13.08       &  0.3358         &     22.40     &  0.5926       &     6.27   &   0.1692      & \textbf{1.22}& \textbf{0.0335} &   4.70         &      0.1360      &       3.92       &   0.1194          \\
      &CC               &     9.05       &  0.2383         &     18.83     &  0.5087       &     5.22   &   0.1382      & \textbf{1.10}& \textbf{0.0308} &   1.79         &      0.0461      &       3.06       &   0.0895   \\
      &pooled           &    15.68       &  0.3837         &     26.20     &  0.6649       &     11.21  &   0.2815      & \textbf{3.66}& \textbf{0.0946} &   11.04        &      0.2454      &      13.54       &   0.3017  \\ \bottomrule
    \end{tabular}
	\end{center}
\end{table*} 

As mentioned in Sect.~\ref{sec:experiments}, one of goals of this work is to analyse traditional texture descriptors for unknown attack detection. To that end, we select the evaluation dataset and assess the detection performance for the adopted texture descriptors by setting up the same parameters reported for the known attack experiment. 

The corresponding results are presented in Tab.~\ref{tab:unknown-attacks}. We can note that the BSIF descriptor attains again the best detection performance for most speech-to-image domain transformations: a D-EER of 4.94\% for PAc, which is close to the one reported by the known attack scenario (i.e., 3.68\%). In addition, the MB-LBP outperforms the remaining descriptors for the STFT-LAc scenario, achieving a D-EER of 10.69\%. 

Based on this fact, we also evaluated the combination between BSIF and FV for each particular PAI species for the LAc and PAc scenarios in Tab.~\ref{tab:fv-unknown-attacks-LA-PA} and established a benchmark against the baselines $B01$ and $B02$. As it can be observed, the CQT achieves the best detection performance for the entire set of LAc-PAIs (i.e., a D-EER of 6.83\%). In addition, this outperforms the adopted baselines for the most challenging PAIs for LAc scenario under the ASVSpoof 2019 challenge~\cite{Todisco-ASVSpoof-arXiv-2019}: D-EERs of 7.91\% and 1.94\% are respectively achieved for A10 and A13, which are two and five times lower than the ones reported by the baselines. Moreover, their corresponding t-DCF values are better than the ones attained by the baselines. 

Consequently, for the Physical Access scenario our CQT-based FV approach attains for the pooled a D-EER of 3.66\%, which is three times lower than the one yielded by the baselines (i.e., a D-EER of 11.04\% for $B01$ and a D-EER of 13.54\% for $B02$). Furthermore, we outperform the baselines for most PAI species: a D-EER in the range 1.10-7.64\% together with a t-DCF between 0.03-0.20\% unveils a reliable and secure generalisation capability for this scenario.  

\begin{figure}[!t]
	\centering
	\includegraphics[width=\linewidth]{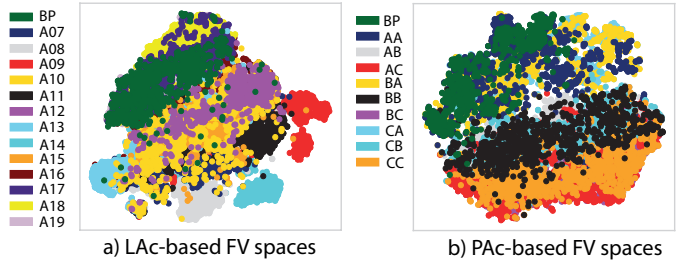}
	\caption{t-SNE visualization of common feature spaces learned by the FV-based approach for the CQT transformation.}
	\label{fig:FV_space}
\end{figure}

Finally, a tSNE visualisation in Fig.~\ref{fig:FV_space} shows that most PAI species share more homogeneous features with each other than with those bona fide presentations. However, the overlap of some PAI species such as A17, A18, AA, BA, and CA with the BP features indicates that the data distribution learned by a GMM using the BSIF features needs to be improved in order to get a better generalisable FV common feature space.     

\subsection{In depth performance analysis}
\label{sec:det}

\begin{figure}[!t]
	\centering
	\includegraphics[width=\linewidth]{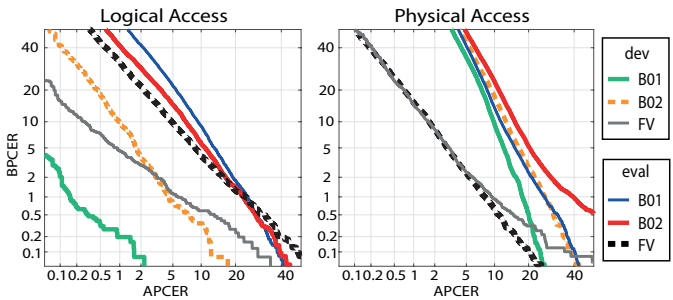}
	\caption{DET curves for the best FV parameter configuration against the baselines for the LAc and PAc scenarios.}
	\label{fig:DET}
\end{figure}

In order to determine whether the FV encoding together with BSIF is suitable to yield a secure ASV system, we evaluated it for several operating points according to the ISO~\cite{ISO-IEC-30107-3-PAD-metrics-170227}. Fig.~\ref{fig:DET} shows the DET curves as well as the BPCER values for several security thresholds for the LAc and PAc scenarios over the development and evaluation datasets. As it should be noted, different behaviours are observed for both the known and unknown attack detection. Whereas the FV encoding outperforms only one out of two baselines approaches in the known attack detection (i.e., dev set) for the LAc scenario (thin gray line), this shows a considerable improvement for the PAc scenario. Specifically, it reports a BPCER100 of 15.24\% which outperforms the baselines results by a relative 81.3\%.\\

On the other hand, for the unknown attack detection (i.e., eval set), we can see that, the FV encoding is better than both baselines: it improves $B01$ and $B02$ by a relative 30.46\% for LAc and 81.67\% for PAc, thereby confirming the results shown in Tab.~\ref{tab:fv-unknown-attacks-LA-PA} and hence its soundness to detect this frequent type of threat. Finally, it is worth noting that the combination of BSIF together with the FV representation reports a similar behaviour for known and unknown attacks over the PAc scenario (i.e., gray vs. black on the right DET figure). In particular, for a high security threshold (i.e., a APCER of 1\%), the FV attains a BCPER value of 15.56\% for unknown attacks, which is marginally worse than the one computed for the known attack detection (i.e., 15.24\%). Based on these results, we confirm that the texture characteristics provided by the CQT domain transformation, captured by the BSIF descriptors, and finally encoded by the FV, yields a secure and convenient ASV system. 

\subsection{Computational efficiency}
\label{sec:efficiency}
In order to evaluate the computational efficiency, we selected the FV encoding for the best parameter configuration (i.e., CQT) and studied its efficiency for several GMM parameter configurations. In the experiments, a PC with a Intel Core i7-8750H processor at 2.2GHz, 16GB RAM was employed. The experimental results report that the FV encoding for $K$ = 512 clusters is able to detect an attack presentation attempt in 1.6 sec approximately for CQT samples with a large size in the range 983-1545 $\times$ 836 pixels. This result indicates in turn that the FV approach is suitable to build a real-time audio PAD system.       

\vspace{-.05cm}

\section{Conclusion}
\label{sec:conclusion}

In this work, we studied the reliability of image texture analysis to develop a secure ASV system. To that end, four traditional texture descriptors which have been widely employed for several PAD tasks were selected. In addition, an application of the well-known FV encoding to audio PAD was proposed. This representation allows the definition of semantic groups given a generative model, such as a GMM.

Experimental results over the ASVSpoof 2019 database confirmed the soundness of the proposal to detect both the known and unknown audio attacks. In particular, the texture characteristics provided by the CQT domain transformation and captured by the BSIF descriptor, reported a D-EER of 16.13\% and 5.97\% for the LAc and PAc scenarios, respectively, in the presence of unknown attacks. These BSIF descriptors represented in turn in a new common feature space by the FV approach achieved a relative performance improvement of 57.66\% for the unknown attack detection, thereby outperforming the current audio-based baselines. In addition, this FV pipeline attained a classification time for large images of 1.6 sec., approximately. Finally, a security analysis of our FV proposal confirmed its generalisation capability: a BPCER100 of 15.56\% in the presence of unknown attacks, which is roughly five times lower than the one attained by the baselines, results in a secure and convenient ASV system.      

As future work research direction, we will investigate a new generative model to tackle the overlapping issues depicted in Fig.~\ref{fig:FV_space}, and hence improve the generalisation capability of the FV representation.
Furthermore, one natural extension of the work would be to look for fusion strategies among methods that are complementary to texture-based PAD.



\vspace{.05cm}




\bibliographystyle{IEEEtran}
\bibliography{2020_WIFS_TextAnalysisforASV_PAD}
%

%
%
%
%
%

\end{document}